\documentstyle[psfig,epsf]{lamuphys}
\makeatletter
\let\chapter\hid@chapter
\makeatother
\def\oo {\Omega_{o}}
\def\Q {Q_{10}}
\def\etal   {{\sl et al.}~\rm}
\titlerunning{Cosmology in a nutshell}

\def\thefirstfig{
\makebox{
\medskip
\noindent
\parbox[l]{1.6truein}{
\footnotesize
{\bf Figure 1. Big Bang vs Steady State.}
Expanding horizon volumes in the Big Bang (A) and
Steady State (B) models. The dots represent the
density of the Universe. At early times only the Big Bang 
model was dense and hot providing an oven for nucleosynthesis
and later an opaque surface of last scattering which naturally 
produces the Planckian spectrum of the CMB.}
\hglue0.3truein
\parbox[r]{2.8truein}{\epsfxsize=2.8truein\epsfbox{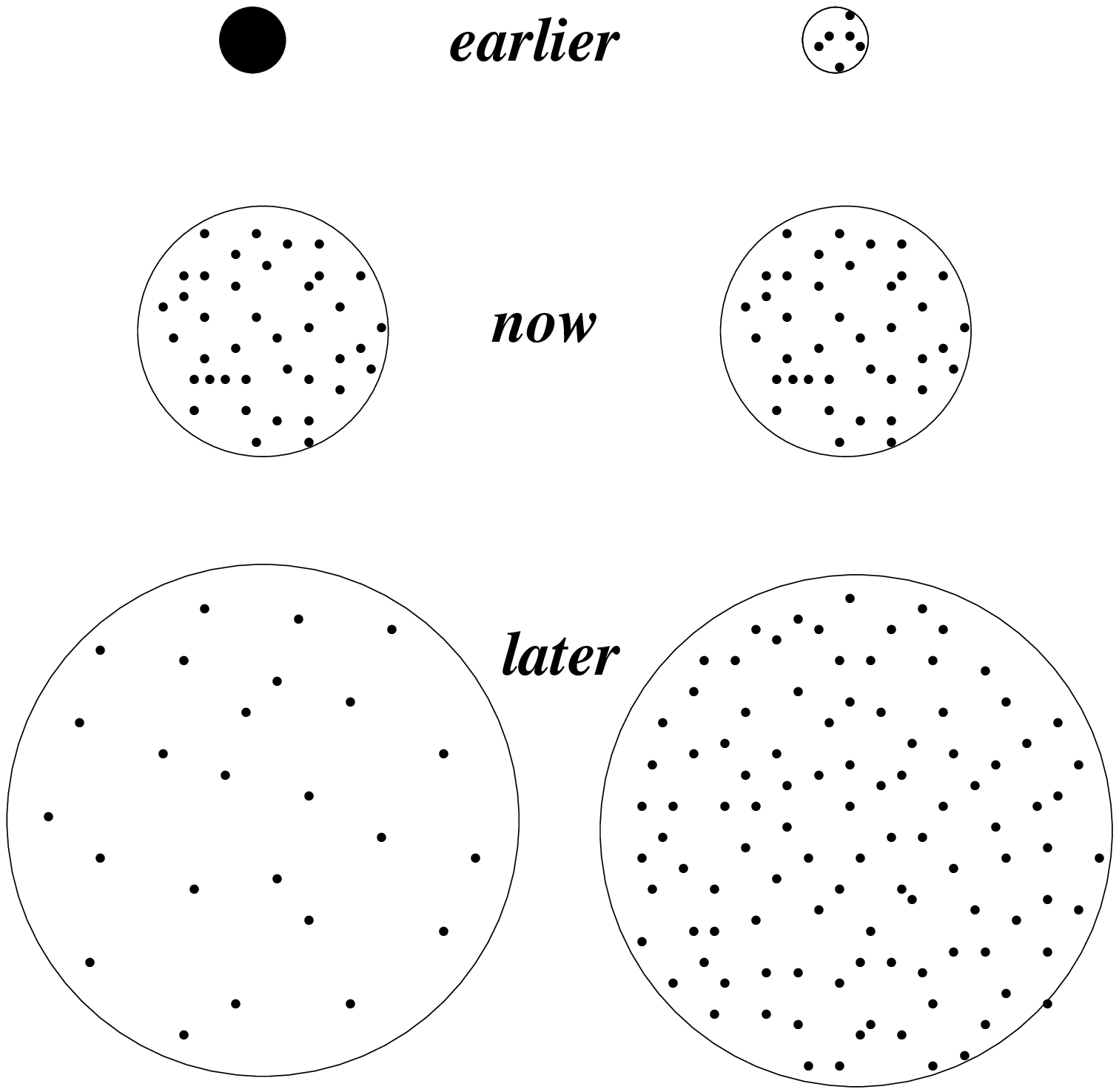}}
\smallskip
}
}

\def\thesecondfig{
\hspace{-0.3cm}
\makebox{
\medskip
\noindent
\parbox[l]{1.6truein}{
\footnotesize
{\bf Figure 2. Gravitational collapse vs topological defects.}
The leading model of structure formation is on the left:
small over-densities are gravitationnally unstable and 
collapse under their own gravity to form the structures
we see around us. Topological defect models (right) are an
alternative. When symmetry is broken in the early universe
causally disconnected regions are occupied by different vacuum states
(indicated by the direction of the lines in the figure).
Large energy densities are present at the boundaries 
between such regions and this is where structure forms.
}
\hglue0.25truein
\parbox[r]{2.6truein}{\epsfxsize=2.6truein\epsfbox{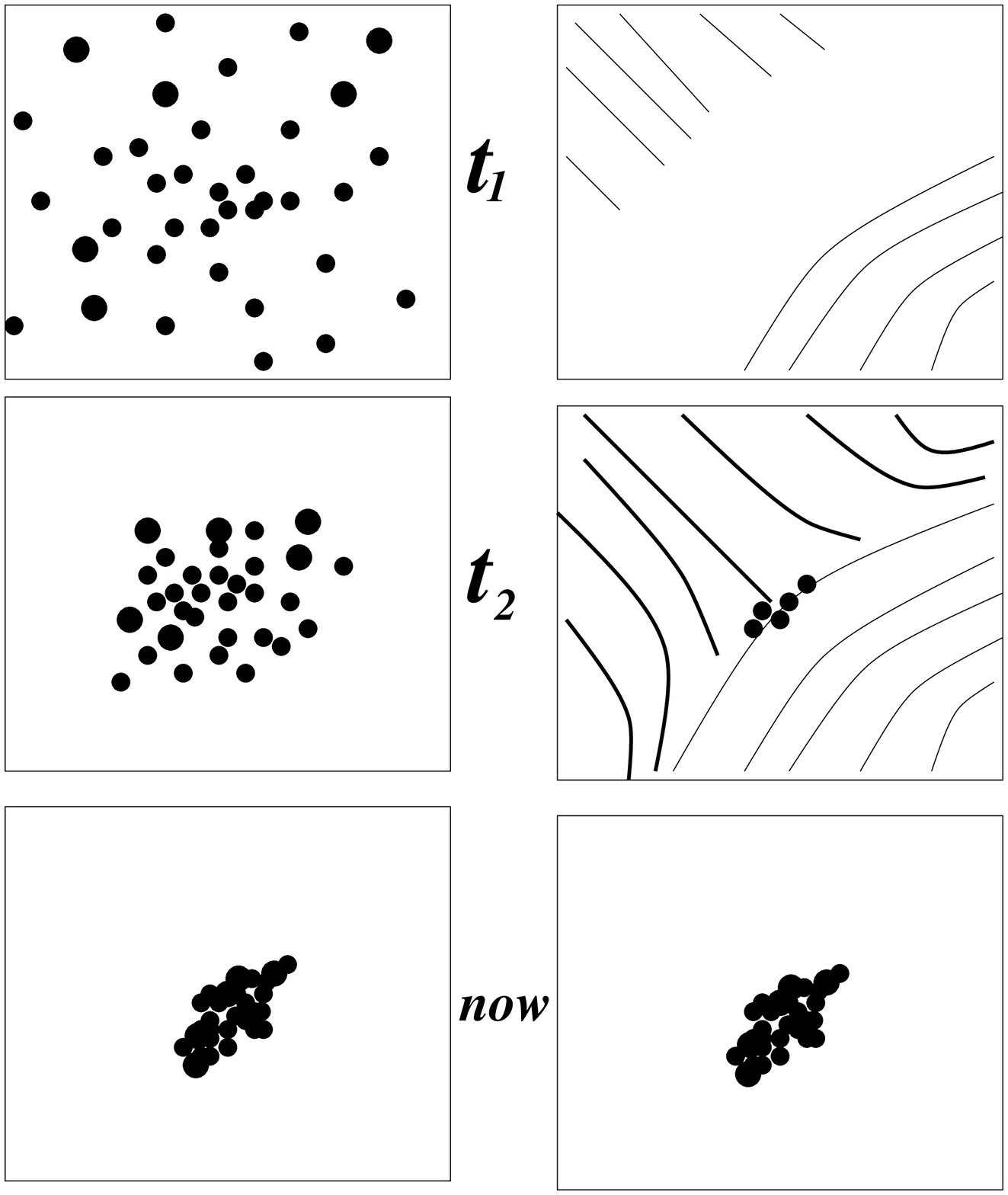}}
\smallskip
}
}

\def\thefourthfig{
\hspace{-0.8cm}
\makebox{
\medskip
\noindent
\parbox[l]{1.74truein}{
\footnotesize
{\bf Figure 4.\\ }
CMB powerspectrum ( $C_{\ell}$, top) compared with the 
matter density powerspectrum ( $P(k)$, middle and bottom). 
$C_{\ell}$ is a measure of the power in the spatial variations of
the CMB as a function of the angular scale.
The upper axes give the angular scales and comoving sizes 
corresponding to the Legendre polynomial index $\ell$.
In the bottom two panels
the turnover in $P(k)$ at $\sim k_{eq}$ occurs
at the scale ($L_{eq} = 2\pi k_{eq}^{-1}$) which just 
enters the horizon as the Universe changes from radiation dominated to matter dominated. Smaller sizes entered the 
horizon earlier during radiation domination and were unable to grow during this period. The $k^{4}$ growth suppression
which they suffered is indicated.}
\hglue0.15truein
\parbox[r]{2.6truein}{\epsfxsize=2.6truein\epsfbox{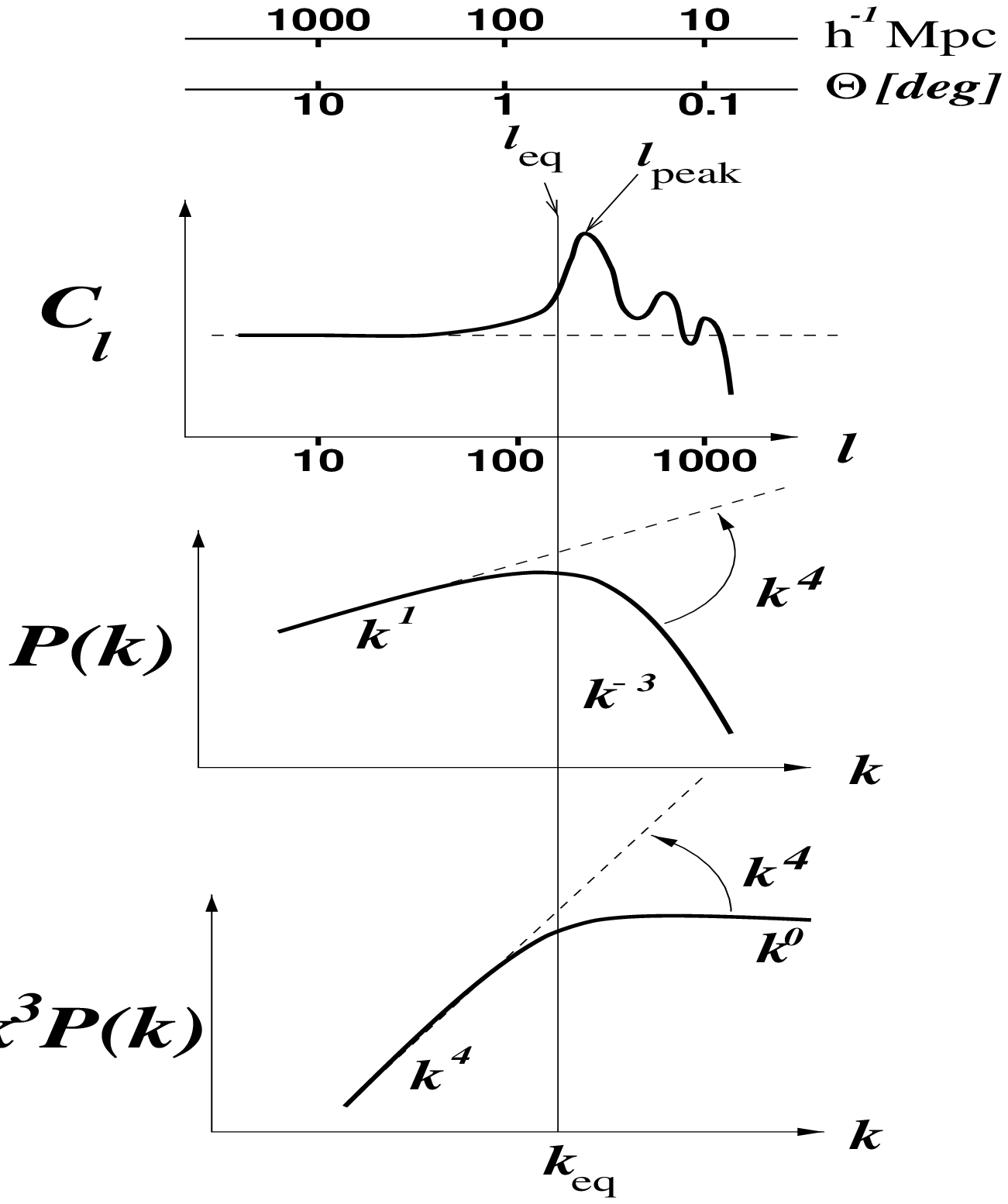}}
\smallskip
}
}

\def\thefifthfig{
\hspace{-0.8cm}
\makebox{
\medskip
\noindent
\parbox[l]{1.7truein}{
\footnotesize
{\bf Figure 5.\\ }
Sound waves in the photon-baryon fluid create bumps in the 
CMB power spectrum.
The grey spots are cold dark matter potential wells which
initiate infall and then oscillation of the photon-baryon fluid 
in these wells.
The Doppler and adiabatic effects make the 
sound visible in the radiation when the baryons decouple from the 
photons during the interval marked $\Delta z_{dec}$.
These bumps are analogous to the standing waves of the resonant frequencies
of a plucked string or of a good shower and  may be the oldest music in 
the Universe.
See Hu {\it etal} (1997) and Lineweaver (1997) for details.}
\hglue0.08truein
\parbox[r]{2.8truein}{\epsfxsize=2.8truein\epsfbox{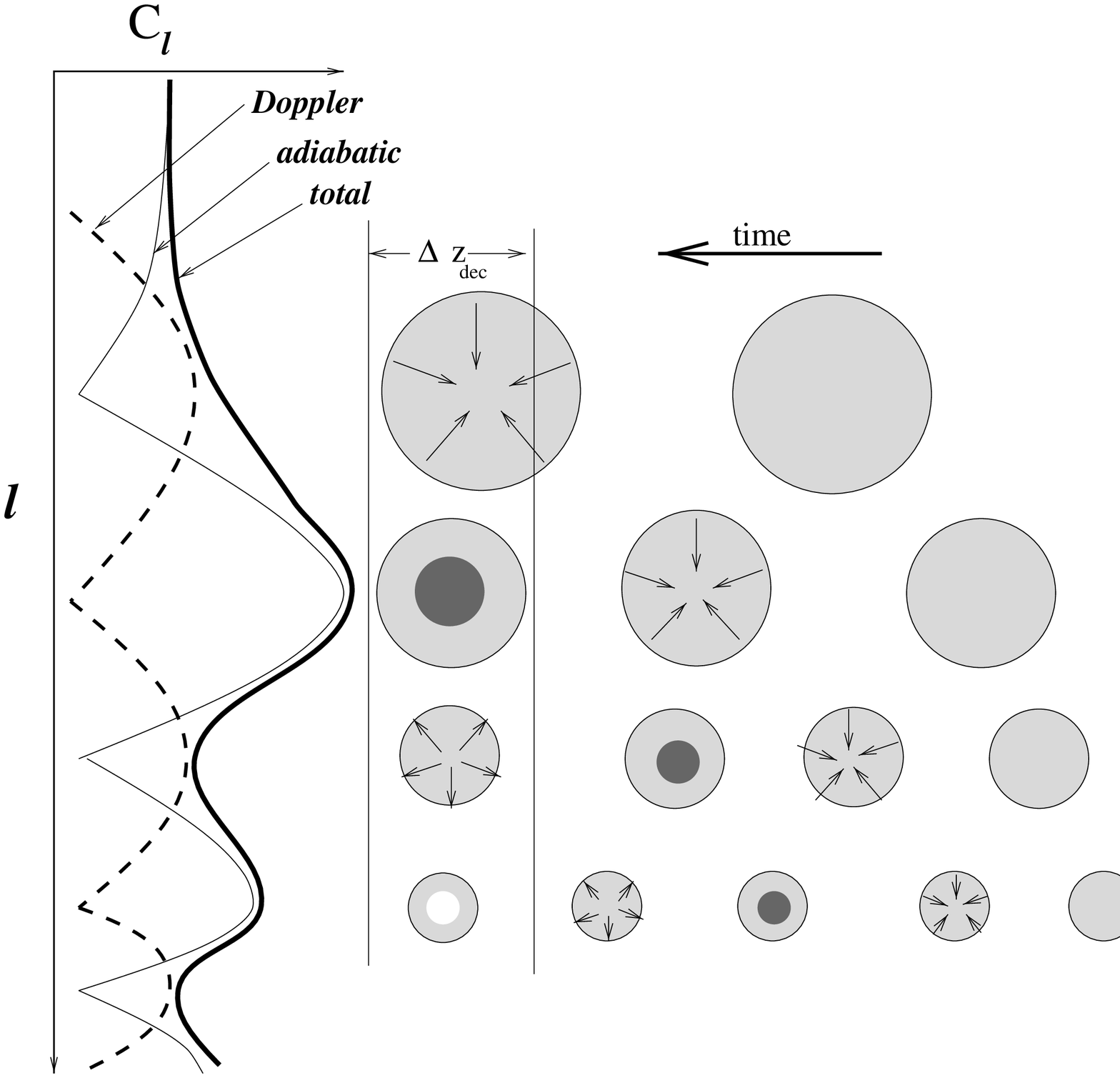}}
\smallskip
}
}

\def\thesixthfig{
\hspace{-0.65cm}
\makebox{
\medskip
\noindent
\parbox[l]{1.7truein}{
\footnotesize
{\bf Figure 6.\\ }
The dark grey banana-shaped region is the approximate
68\% confidence level preferred by the current CMB anisotropy
measurements. The thick solid lines are the approximate
2, 3 and 4 $\sigma$ contours. The age interval shown
is $10 - 18$ Gyr. The thin lines are contours of the spectral index
$n$ values which minimize the $\chi^{2}$ for each pair $(h, \oo)$.
Note the monotonic relations: the higher the $h$ value the
higher the $n$ value and the lower the $\oo$ value.
A favored open model $h \approx 0.70$ with
$\oo \approx 0.3$ is rejected at greater than $\sim 4 \sigma$.
 The best-fit value is
$h=0.40$ and $\oo = 0.85$. The corresponding $n$ value is
$0.91$.Figure adapted from Lineweaver \& Barbosa (1998).}
\hglue0.05truein
\parbox[r]{3.3truein}{\epsfxsize=3.3truein\epsfbox{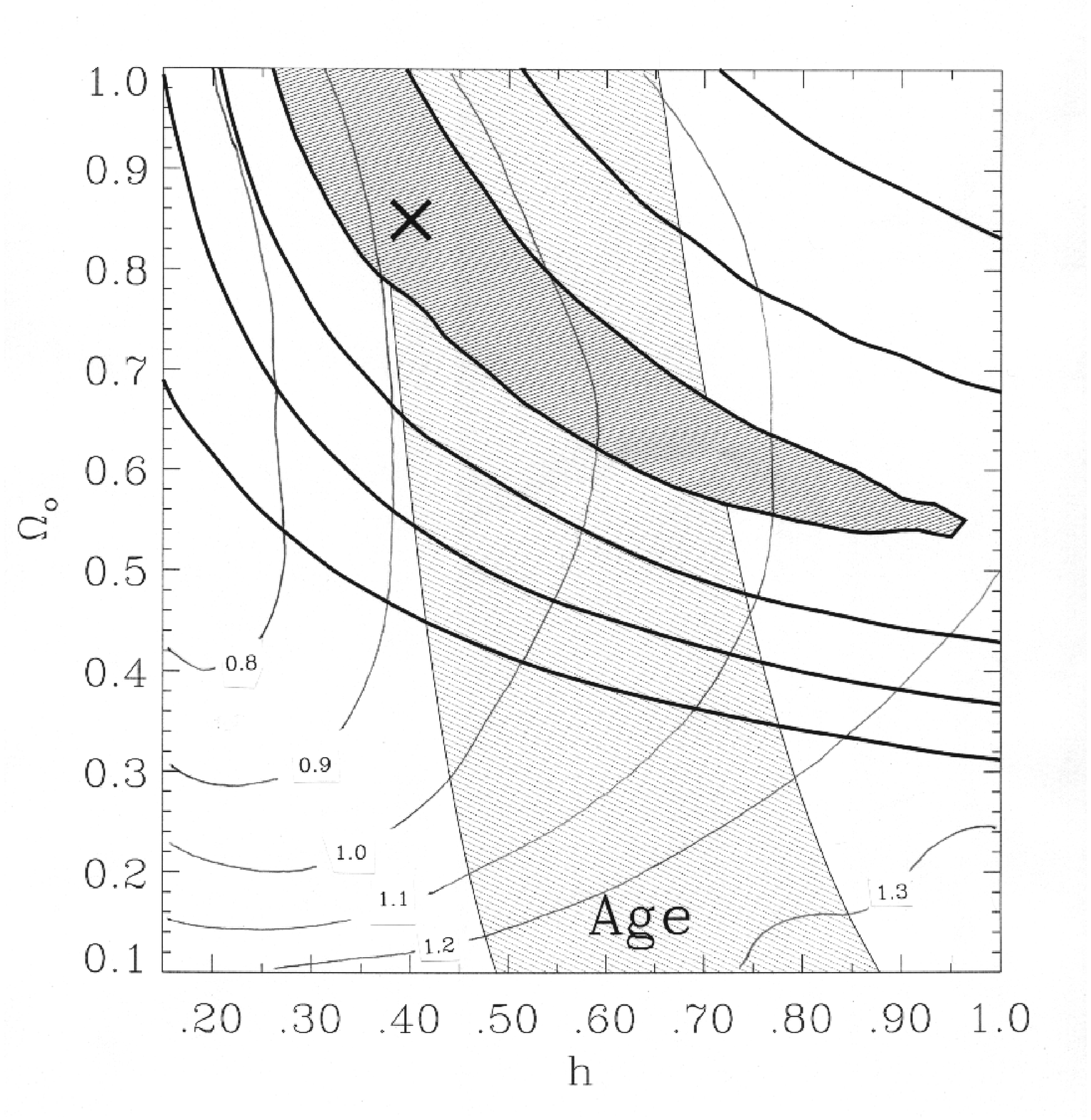}}
\bigskip
}
}

\def\theseventhfig{
\hspace{-0.65cm}
\makebox{
\bigskip
\noindent
\parbox[l]{1.7truein}{
\footnotesize
{\bf Figure 7.\\ }
Contours in the plane of the slope $n$ and normalization
$\Q$ of the primordial power spectrum of CMB anisotropies.
Notation is analogous to the previous figure except here
the thin lines are contours of the $h$ values which 
minimize the $\chi^{2}$ for a given pair of $(n,\Q)$.
Note the monotonic relation: for a given $\Q$ value,
the higher the $n$ value the
higher the $h$ value.
The best-fit values of $n$ and $\Q$ are
$n = 0.91^{+0.29}_{-0.09}$ and $\Q = 18.0^{+1.2}_{-1.5}$.
To my knowledge, these are the tightest constraints on 
these parameters.
Figure adapted from Lineweaver \& Barbosa (1998).}
\hglue0.05truein
\parbox[r]{3.3truein}{\epsfxsize=3.3truein\epsfbox{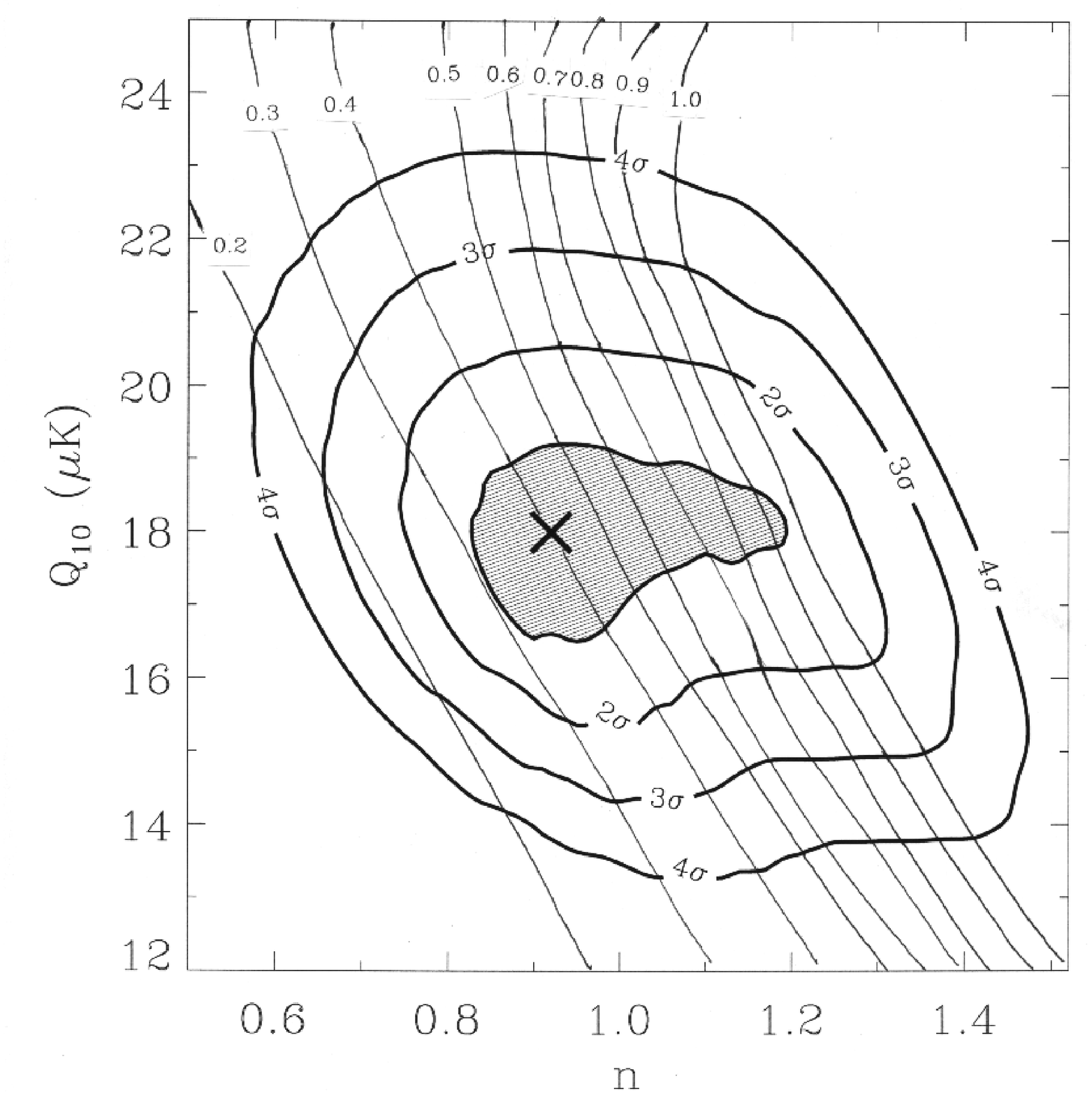}}
\smallskip
}
}

\begin{document}
\pagenumbering{arabic}
\title{Cosmology in a nutshell + an argument against 
$\Omega_{\Lambda} = 0$ based on the
inconsistency of the CMB and supernovae results}
\author{Charles H. Lineweaver}
\institute{University of New South Wales, Sydney, Australia}
\maketitle

\begin{abstract}
I present several simple figures to illustrate cosmology and structure 
formation in a nutshell. Then I discuss the following argument:
if we assume that $\Omega_{\Lambda} = 0$ then the
CMB results favor high $\Omega_{m}$ while the supernova results favor
low $\Omega_{m}$. This large inconsistency
is strong evidence for the incorrectness of the 
$\Omega_{\Lambda}=0$  assumption.
Finally I discuss recent CMB results on the slope and normalization 
of the primordial power spectrum.
\end{abstract}
\section{Cosmology in a nutshell}

The Big Bang model became the standard cosmological model
soon after the discovery of the cosmic microwave background (CMB). 
The Big Bang model has a hot, dense early epoch
(see Figure 1) when nucleosynthesis occurred. It also has
an opaque surface that can naturally produce the Planckian spectrum 
of the CMB. The Steady State universe was not 
hotter in the past, has no epoch of Steady State Nucleosynthesis
and has no opaque surface to produce the CMB.

Gravitational collapse is the leading model of structure 
formation (Figure 2).
Slight over-densities are gravitationally unstable and collapse
under their own self-gravity. 
In an alternative family of models, structure forms from topological defects.
In gravitational collapse models 
CMB anisotropies larger than $\sim 1$ degree are acausal and 
rely on inflation to explain their existence. In defect models
these large anisotropies are close by, causal and sub-horizon sized.
\thefirstfig

\clearpage
\thesecondfig

\setcounter{figure}{2}
\begin{figure}[hbt]
\centerline{\psfig{file=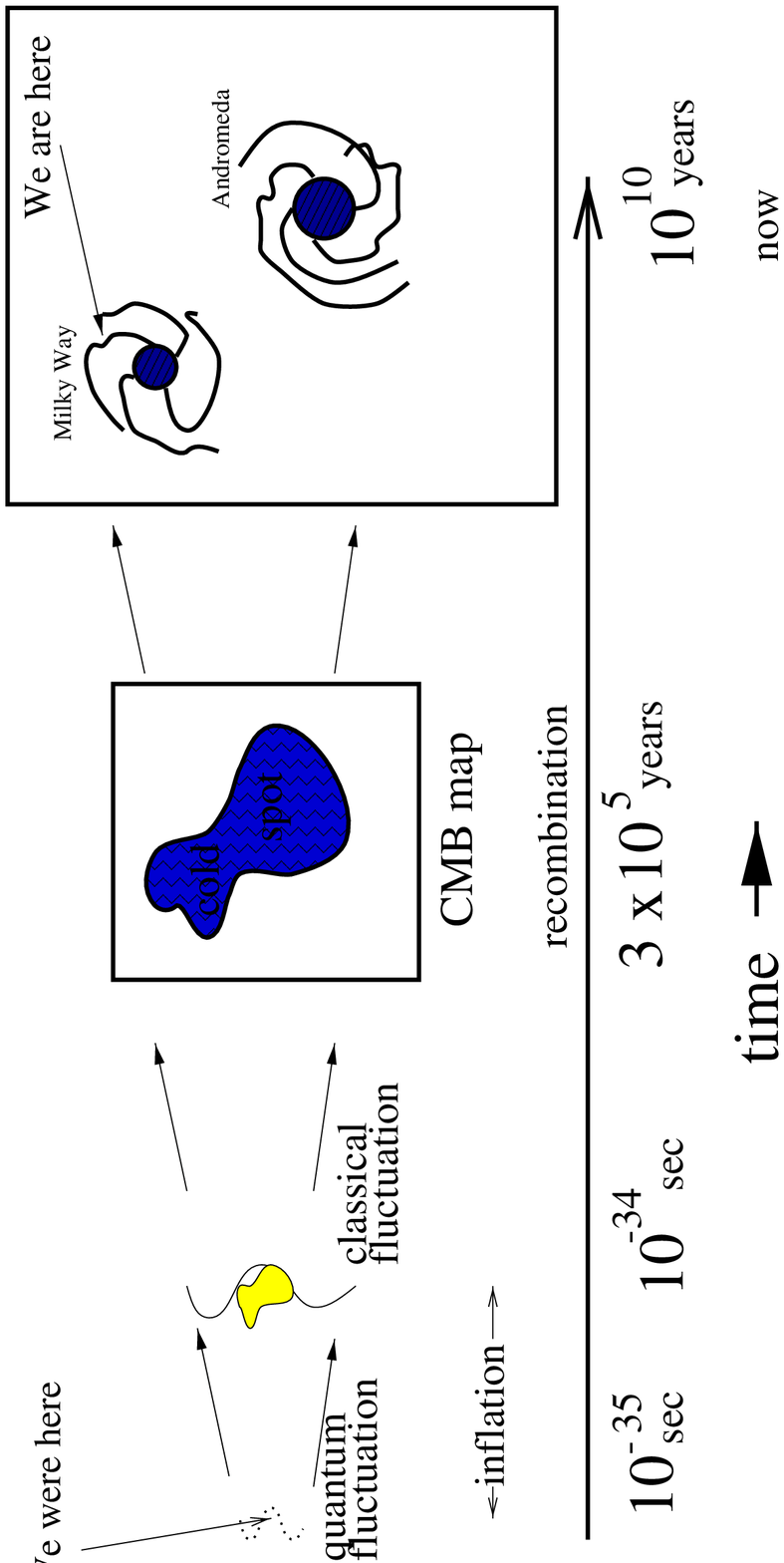,height=70mm,width=110mm,angle=-90}}
\vspace{0cm}
\caption{Galaxies are CMB anisotropies are Quantum fluctuations.
According to the inflationary scenario, quantum fluctuations of a
scalar field are the origin of all structures.
These quantum fluctuations are not caused by any preceeding event
in the same sense as radioactive
decay or quantum tunneling are not caused.
They are non-deterministic prime movers.
Inflation of the universe by a factor of more than $10^{26}$ 
transforms these quantum fluctuations into super-horizon classical density fluctuations. On their way to becoming galaxies we can monitor their 
progress by looking at CMB maps.}
\end{figure}

\clearpage
One of the most important questions in cosmology is:
What is the origin of all the galaxies, clusters, great walls, 
filaments and voids we see around us?
The inflationary scenario provides the most popular 
explanation for the origin of these structures: 
they used to be quantum fluctuations.

Figure 3 illustrates the metamorphosis of quantum fluctuations
to CMB anisotropies to galaxies. Primordial quantum fluctuations 
of a scalar field get amplified and evolve
to become classical seed perturbations and eventually
large scale structure.
This process can be monitored by CMB observations since matter
fluctuations produce temperature fluctuations in the CMB:
$\frac{\delta \rho}{\rho} \propto \frac{\Delta T}{T}$.

How does a particular fluctuation know whether it will become
a spiral or an elliptical galaxy?
Does the density and irregularity of its environment determine its
morphology by controlling its angular momentum and the amount of merging?
With a full understanding of galaxy formation we may be able to look 
at CMB cold spots and their neighborhoods and predict where they will 
end up in the Hubble tuning fork diagram of galaxy types.
The distribution of morphological types
at high redshift discussed by Driver in these proceedings
would then be a derivable function of the characteristics
of the CMB anisotropies.

\thefourthfig

\subsection{There is no scale beyond which the universe is homogeneous}

It has been claimed that some recent, deep, galaxy redshift surveys 
have reached the scale at which the Universe becomes homogeneous.
Strictly speaking however there is no scale beyond which the universe is
homogeneous. The amplitude of
the density contrast ($\delta \rho / \rho \propto k^{3}P(k)$ 
decreases for larger scales but is never zero.
A more meaningful question is: Where is the turnover
in the power spectrum?
This turnover is due to a suppression of growth of a 
given k mode by $k^{4}$ relative to modes which enter the
horizon during matter domination (assuming $\oo = 1$).
Thus, the horizon scale at matter-radiation equality
is an important diagnostic of this fundamental scale.
See Figure 4. 


Lineweaver \& Barbosa (1998) have used current CMB
anisotropy measurements to determine
the position of the adiabatic peak in the CMB spectrum under
the assumption of open or critical density CDM dominated 
universes: $\ell_{peak} = 260^{+30}_{-20}$.

Figure 5 illustrates how harmonic sound bumps appear in the CMB power spectrum driven by the wells and valleys of the CDM 
potentials.
The epoch when matter and radiation densities are equal has a redshift of $z_{eq}$ while
decoupling occurs at $z_{dec}$.
 The number of oscillations between $z_{eq}$ and $z_{dec}$ and thus the phase of the oscillations at $z_{dec}$ is determined by i) the physical size of the potential well, ii) the speed of sound and iii) the time interval between $z_{eq}$ and $z_{dec}$. 

\vspace{0.7cm}

\thefifthfig
\clearpage

\section{ An argument against $\Omega_{\Lambda} = 0$ based on
the inconsistency of the latest CMB and supernovae constraints
on $\Omega_{m}$.}

The CMB is already giving us 
useful constraints on cosmological parameters
in popular but restricted families of CDM models.
In Figure 6 the region of the $h-\Omega_{m}$ plane
preferred by the CMB data is shown
(since $\Omega_{\Lambda}=0$, $\Omega_{0} = \Omega_{m}$).
The best-fit is indicated
with an {\bf X}. The values of the spectral index $n$ which minimize the $\chi^{2}$
values for a given $(h,\Omega_{o})$ pair are indicated by
the thin solid iso-n lines and are labeled with $0.8-1.3$. 
High values of $h$ require high values of $n$.

\vspace{0.2cm}
\thesixthfig
\vspace{0.2cm}

Under the assumption that $\Omega_{\Lambda}=0$ 
(i.e., $\Omega_{o}= \Omega_{m}$) the most recent
CMB results on the density of matter in the Universe yield
$\Omega_{m} > 0.3$ at the $\sim 4\sigma$ confidence level 
(Figure 6).
This result is independent of the value of Hubble's constant,
of the spectral index $n$ and of the normalization $\Q$. 
In this same model the new supernovae results 
are $\Omega_{m} = -0.1 \pm 0.5$ (Garnavich \etal 1998)
and $\Omega_{m} = -0.4 \pm 0.1$ (statistical) $\pm 0.5$ (systematic)
(Perlmutter \etal 1998).
With additional supernovae $\Omega_{m}$ stays low and 
the error bars decrease making the inconsistency
between CMB and supernovae results even
stronger (Schmidt 1998 private communication).
People who like open models with $\Omega_{\Lambda} = 0$
could argue that these supernovae results are consistent
with some cosmological measurements which yield 
$0.1 \lse \Omega_{m} \lse 0.3$.
However the strong inconsistency between the CMB results
(which strongly exclude low values of $\Omega_{m}$) 
and the supernovae which favor very low, even negative
(and thus unphysical) values of $\Omega_{m}$ is
strong evidence against $\Omega_{\Lambda} = 0$ models.

\section{Results for the slope and normalization of the CMB power
spectrum}
The CMB solutions for $h$ or $n$ can be read from either
Figure 6 or 7. 
CMB anisotropy measurements in open and critical
CDM models yields  $n= 0.91^{+.29}_{-0.09}$ (Lineweaver \& Barbosa 1998).
In Figure 7 the thin solid lines indicate the 
$h$ values ($0.2 \le h \le 1.0)$ which minimize
the $\chi^{2}$ for $(n,\Q)$ pairs.
Conditioning on $n=1$ or $h=0.50$ changes the results
(see Table 2 of Lineweaver \& Barbosa 1998).

\theseventhfig

%

\end{document}